\documentclass{iau}
\usepackage{graphicx}

\title[Large Spectroscopic Surveys] 
{Studying Young Stars with Large Spectroscopic Surveys}

\author[S.~L.~Martell]   
{Sarah L.~Martell$^1$}

\affiliation{$^1$School of Physics, University of New South Wales\\email: {\tt s.martell@unsw.edu.au}}

\pubyear{2015}
\volume{314}  
\pagerange{119--126}
\jname{Young Stars \& Planets Near the Sun}
\editors{J. H. Kastner, B. Stelzer, \& S. A. Metchev, eds.}
\begin{document}

\maketitle

\begin{abstract}
Galactic archaeology is the study of the history of star formation and chemical evolution in the Milky Way, based on present-day stellar populations. Studies of young stars are a key anchor point for Galactic archaeology, since quantities like the initial mass function and the star formation rate can be studied directly in young clusters and star forming regions. Conversely, massive spectroscopic Galactic archaeology surveys can be used as a data source for young star studies.
\end{abstract}

\firstsection 
\section{Introduction}

The goals of Galactic archaeology, as laid out by \cite[Freeman \& Bland-Hawthorn (2002)]{FBH02}, are to identify stars that formed together using their surface abundance patterns, and then to use those groups of co-formed stars to explore the history of star formation and chemical enrichment in the Milky Way. Since the typical mode for star formation is clustered (e.g., \cite[Lada \& Lada 2003]{LL03}), and star-forming regions up to the masses of globular clusters ($\sim$ few $\times 10^{5} M_{\odot}$) ought to be chemically well-mixed (e.g., \cite[Feng \& Krumholz 2014]{FK14}), we expect that each star formation event should produce a large number of stars with matching abundance patterns.

During these stars' lifetimes they will drift, scatter and migrate through the Galaxy. This process begins with the evaporation of young clusters (e.g., \cite[Terlevich 1987]{T87}; \cite[Murphy et al. 2010]{MLB10}). However, the majority of a star's surface abundances will remain constant as it evolves, and it is this persistence of the initial abundance pattern that allows stars from those initial groups to be ``chemically tagged'' as belonging together (e.g., \cite[De Silva et al. 2011]{DS11}; \cite[2013]{DS13}). 

Identifying stars with matching abundance patterns is an interesting challenge - because some elements are produced in multiple nucleosynthetic sites, on different timescales, there is not a simple model for how stars should be arranged in the parameter space of chemical abundances (``C-space''). Simply matching as many abundances as can be measured is inefficient since there are a number of element groups that tend to be produced in the same astrophysical sites and enriched or depleted simultaneously. A number of ongoing and planned survey projects will be determining a large number of elemental abundances per star for many stars; quantifying the most efficient and informative ways to search for groups within that abundance data will be an important challenge for those survey teams.

\section{Spectroscopic Galactic archaeology surveys}
Chemical tagging can be carried out at different levels with the data from many large survey projects, whether it is focused on searching for unusual stars (e.g., \cite[Martell \& Grebel 2010]{MG10}; \cite[Martig et al. 2014]{MR14}), or on identifying groups of stars that formed together (\cite[Bland-Hawthorn et al. 2010]{BH10}; \cite[Mitschang et al. 2013]{MD13}). The abundance precision needed to select co-formed groups of stars is quite high, estimated as $0.05$ to $0.1$ dex in [X/H]\footnote{By convention, elemental abundance ratios written in square brackets are expressed on a logarithmic scale normalised to the Sun. For an element X in a star, [X/H]$={\rm log(N_{\rm X}/N_{\rm H})_{\rm star}}-{\rm log(N_{\rm X}/N_{\rm H})_{\rm \odot}}$}. This is a difficult but achievable level of precision, even for large survey projects with automated data reduction and analysis processes.

Recent theoretical work evaluating the prospects of identifying co-formed stars based solely on abundance information (\cite[Ting et al. 2015]{TC15}) has found that it may be quite difficult, depending on the intrinsic differences between the abundance patterns of independent groups of stars. If there is not much separation between the groups in C-space, then it will naturally be quite difficult to distinguish them. However, the addition of age information (for example, requiring that stars with similar abundance patterns must also fall near a single isochrone, as is done in \cite[Mitschang et al. 2014]{MD14}) is a powerful tool in confirming that a group of stars did form at the same time. Even in the case when individual co-formed groups cannot be identified, the form and strength of structure in C-space still carries information about the star formation rate, the initial cluster mass function, and the minimum mass for star-forming events.

Since Galactic archaeology is so interested in the mode and rate of star formation, studies of present-day young stars have the potential to be quite important to our understanding of Galactic history. As one example, direct measurements of the stellar initial mass function in a young cluster or moving group are far more concrete than distributions inferred from much older populations. In addition, detailed abundance determinations for young stars can be used to test whether star-forming regions are as well-mixed as predicted, and how well the abundance patterns of co-forming stars match. 

Ironically, the level of detail available to studies of present-day young stars can make it difficult to combine them with old samples. The ages for old stars typically carry error bars of $\pm 1$ Gyr, and studies based on old stars would not be able to distinguish substructures within a star-formation region where the stars formed $10-100$ Myr apart. Studies of pre-main-sequence stars can determine ages to a much finer resolution, and as a result the membership criteria for young associations and moving groups (e.g., \cite[Malo et al. 2013]{MDL13}) can be quite complex. The processes and algorithms for group-finding in Galactic archaeology data will need to recognise the fact that star formation is a distributed and asynchronous process.

\section{Opportunities for young star studies in existing and future survey data sets}
The three large high-resolution spectroscopic Galactic archaeology surveys currently in operation (GALAH, \cite[de Silva et al. 2015]{DF15}; APOGEE, \cite[Holtzman et al. 2015]{HS15}; Gaia-ESO, \cite[Gilmore et al. 2012]{G12}) all take different approaches to including young stars in their sample. Gaia-ESO specifically observed a number of young associations, which have produced some of their first scientific results. \cite[Spina et al. (2014)]{SR14} find that the metallicity in the Chamaeleon 1 star-forming region is mildly sub-Solar, similar to other star-forming regions in the Solar neighbourhood. \cite[Jeffries et al. (2014)]{JJ14} find that the $\gamma$ Velorum cluster comprises two independent groups along the same line of sight, with slightly different systemic velocities and ages.

APOGEE devotes a fraction of its observing time to ancillary science, and the IN-SYNC ancillary program (\cite[Cottaar et al. 2014]{CC14}) has observed nearly 3500 young ($\leq 10$ Myr) stars across IC 348 and the Pleiades. They find that there is a range in radius at fixed stellar mass in IC 348, indicating an age range of $25\%$. GALAH has a small number of specifically targeted fields, including the Kepler-2 (K2, \cite[Howell et al. 2014]{HS14}) Campaign 2 field, which covers the Scorpius-Centaurus association. The targeting in GALAH is aimed at red giants to be observed by K2, and not specifically at young stars, but nevertheless a useful fraction of the over $2000$ stars reported as candidate young stars by \cite[Rizzuto et al. (2015)]{RI15} have been observed by GALAH. Given the very thorough sky coverage of GALAH, it is quite likely that other star-forming regions, young associations and young clusters will also be part of the GALAH sample.

Finally, there are other spectroscopic resources that bear exploration and data-mining. The WEAVE (\cite[Dalton et al. 2012]{DT12}) and 4MOST (\cite[de Jong et al. 2012]{DJ12}) surveys, which are both due to start observing in the next few years, will collect massive data sets and could involve targeted observations of young stars, with involvement from the community. The Large Sky Area Multi-Object Fibre Spectroscopic Telescope (LAMOST, \cite[Luo et al. 2015]{LZ15}) survey has been acquiring low-resolution (R$\sim 1800$) spectra since 2011, and its first data release includes spectra and stellar parameters for over one million A, F, G and K stars. The first LAMOST data release also includes over 120,000 stars classified as M dwarfs based on their spectra, providing a rich data set in which to search for spectroscopic youth indicators like chromospheric activity and strong Li 6708{\hbox{\AA}} absorption. Figure 1 shows an example of a previously unreported Li-rich M star in LAMOST. This spectrum was identified during a cursory examination of the public LAMOST data, using the Li 6708$\hbox{\AA}$ absorption index defined in \cite[Martell \& Shetrone (2013)]{MS13}. It is classified by LAMOST as an M0 star.

\begin{figure}[b]
\begin{center}
\includegraphics[width=3.4in]{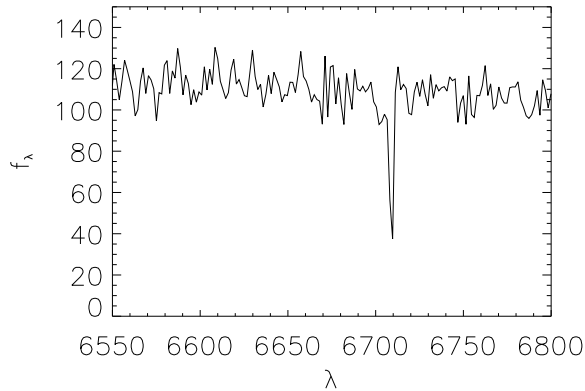} 
 \caption{Portion of a LAMOST spectrum of a previously unknown Li-rich M0 star. Further data mining should uncover more stars like this.}
   \label{fig:li}
\end{center}
\end{figure}

\begin{discussion}

\discuss{Prof. Trimble}{Are there any likely candidates for stars that co-formed with the Sun?}

\discuss{Dr. Martell}{The open cluster M67 has been suggested as a place where the Sun may have formed, but mainly based on age and metallicity. You would need to match the abundance patterns across more elemental abundances to be more certain. Trying to match the Sun to the stars it co-formed with based on Galactic orbits is less certain than chemical tagging, since possibility of scattering by spiral arms or giant molecular clouds in the time since the Sun formed is not negligible.}

\discuss{Prof. Mamajek}{Is anyone already searching the LAMOST data for active stars?}

\discuss{Dr. Martell}{Not that I know of.}

\end{discussion}

\end{document}